\documentclass[runningheads]{llncs}
\usepackage[T1]{fontenc}
\usepackage{graphicx}
\usepackage{amsmath,amssymb}

\usepackage{booktabs}
\usepackage{multirow}
\usepackage{xcolor}
\usepackage{subcaption}
\usepackage{marvosym}
\usepackage{hyperref}

\begin{document}

\title{Leptomeningeal Collateral Detection on DSA via Vessel-Graph Neural Networks}

\titlerunning{Node-Level Collateral Detection via Vessel-Graph GNN}



\author{Junyong Cao\inst{1}$^{\dagger}$ \and
Hakim Baazaoui\inst{2}$^{\dagger}$ \and
Chinmay Prabhakar\inst{1} \and
Suprosanna Shit\inst{1} \and
Lukas Bastian Otto\inst{2} \and
Susanne Wegener\inst{2} \and
Bjoern Menze\inst{1}$^{\star}$ \and
Ezequiel de la Rosa\inst{1}$^{\star}$\textsuperscript{\Letter}}

\authorrunning{J. Cao et al.}
\institute{University of Zurich, Zurich, Switzerland \and
University Hospital Zurich, Zurich, Switzerland\\[0.6em]
$^{\dagger}$$^{\star}$~These authors contributed equally to this work.}

\maketitle
\renewcommand\thefootnote{}%
\footnotetext{\raggedright Code available at \url{https://github.com/JCTaylor666/leptomeningeal-collateral-detection}.}%
\footnotetext{\raggedright \Letter~Corresponding author: \email{ezequiel.delarosa@uzh.ch}}%
\renewcommand\thefootnote{\arabic{footnote}}%

\begin{abstract}

Leptomeningeal collaterals (LMCs) are an important prognostic factor in acute ischemic stroke. Existing automated methods rely on CT angiography (CTA), but individual LMCs are often too small to be resolved on CTA, limiting these methods to coarse collateral scoring. Digital subtraction angiography (DSA) visualizes individual collaterals at superior resolution, yet current assessment remains subjective, relying on manual grading scales that suffer from poor inter-rater agreement. We present a framework that formulates collateral detection as the classification of individual vessel segments on a graph derived from DSA. A hybrid \emph{graph-pixel} architecture combines a topology-aware graph branch with a dense pixel branch, fused in a shared node-probability space. In a five-fold cross-validation setting, the fused model achieves a PR-AUC of 0.434, outperforming the graph-only (0.403) and pixel-only (0.362) baselines. To our knowledge, this is the first method to enable the individualization of LMCs in DSA, allowing for precise per-vessel quantitative assessment. This integration shifts DSA assessment toward objective evaluation, supporting future biomarker and pattern discovery for individual LMCs.


\end{abstract}

\section{Introduction}

Ischemic stroke is a leading cause of death and adult disability worldwide~\cite{campbell2019}, and leptomeningeal collaterals (LMCs) are a key determinant of tissue survival after stroke onset~\cite{liebeskind2003,shuaib2011}. LMCs are pial arteries linking the distal branches of the anterior, middle, and posterior cerebral artery (ACA, MCA, PCA) territories, providing retrograde perfusion to ischemic territory when proximal vessels are obstructed. Robust collateral supply is associated with a larger volume of potentially salvageable tissue (the "penumbra") and better functional outcome; conversely, poor collaterals predict continued infarct growth and treatment complications~\cite{campbell2013,maas2009,mohamed2023}.

Despite their prognostic importance, collateral assessment remains largely rudimentary in clinical practice, if performed at all~\cite{hamann2021}. Pre-treatment modalities such as CT angiography (CTA) and CT perfusion (CTP) are widely used for acute diagnosis and treatment decisions, but their spatial resolution is insufficient to resolve the fine pial network of individual LMCs~\cite{liebeskind2003,shuaib2011}, limiting current assessment to coarse global or regional collateral status.
Digital subtraction angiography (DSA) is regarded as the gold standard for collateral visualization~\cite{christoforidis2005}, providing the highest spatial resolution among clinical imaging modalities and enabling direct visualization of individual collateral branches. However, DSA interpretation demands deep neurointerventional expertise, and manual grading scales such as the American Society of Interventional and Therapeutic Neuroradiology/Society of Interventional Radiology (ASITN/SIR) are subjective, suffering from poor inter-rater agreement even among expert readers~\cite{benhassen2018}.
Automated analysis on DSA has remained largely unexplored, owing to its procedural complexity, limited data availability and the absence of structured vascular representations suitable for learning. Existing automated approaches therefore predominantly target CTA-based global collateral scoring~\cite{kuang2023cta,lu2025u2net}, producing a single scalar per patient that cannot localize individual collateral vessels.
Although pixel-level vessel segmentation methods exist for DSA~\cite{zhang2020dsaseg}, they delineate the vascular tree as a whole, cannot identify which segments are collaterals, and do not explicitly encode vascular topology such as segment connectivity and hierarchical position. This topology is precisely what clinicians use to assess collateral status on DSA, yet no existing method operates at the individual vessel-segment level.

To bridge this gap we present, to our knowledge, the first framework for automated identification of individual LMCs on DSA, moving collateral assessment from subjective ordinal grading toward objective per-vessel scoring. Our key idea is to represent the visible vasculature as a structured graph of connected vessel segments, casting collateral detection as a node-level classification task and mirroring how clinicians reason about collateral pathways along the vascular tree. We further propose a hybrid framework comprising two complementary branches: a graph branch that reasons over vascular topology to capture spatially dispersed collateral patterns, and a pixel branch that provides dense local confirmation. Both branches are projected into a shared node-probability space, enabling unified fusion that consistently outperforms either branch alone. In summary, our contributions are as follows:
\begin{enumerate}
    \item \textbf{Individual-level collateral identification on DSA.} We formulate collateral detection as node-level classification on a vessel-segment graph derived from DSA, enabling quantitative per-vessel scoring.
    \item \textbf{Hybrid graph-pixel framework.} We introduce a unified model that combines a graph-based node detector with a pixel-based branch in a shared node-probability space, allowing topology-aware reasoning and dense local evidence to be modeled jointly.
    \item \textbf{Complementary analysis and fusion in shared node space.} We show that graph and pixel branches capture complementary collateral patterns, and that their fusion improves overall performance beyond either branch alone.
\end{enumerate}

\section{Related Work}
\paragraph{Automated collateral assessment.}
The vast majority of automated collateral evaluation methods operate on CT angiography (CTA), multi-phase CTA or CT perfusion (CTP). These approaches use deep-learning models to predict a patient-level or region-level collateral score~\cite{kuang2023cta,lu2025u2net,lin2020perfusion}. These approaches produce a single global scalar per patient and cannot localize individual collateral vessel segments. Moreover, CTA and CTP lack the spatial resolution to resolve individual pial collaterals, and CTA- and DSA-based assessments are not directly interchangeable~\cite{shuaib2011}. To date, no prior work has attempted automated detection of individual collateral vessels on DSA.

\paragraph{DSA vessel segmentation and graph construction.}
Neural-network-based vessel segmentation on DSA has been shown to be feasible using encoder--decoder architectures~\cite{zhang2020dsaseg}, with nnUNet~\cite{isensee2021nnunet} providing a strong self-configuring baseline. However, pixel-level masks alone cannot express vascular topology. Several works convert segmentation masks into graph representations via skeletonization and junction detection~\cite{chapman2015vesselgraph}. In the vasculature domain, the VesselGraph benchmark~\cite{vesselgraph2022} formalizes a line-graph convention in which vessel segments become graph nodes and shared junctions become edges. We adopt this convention to bridge the gap between dense segmentation and topology-aware analysis.

\paragraph{Graph neural networks for anatomical structures.}
Graph neural networks (GNNs) have been increasingly applied to node-level tasks on anatomical tree structures~\cite{mienye2025gnnsurvey}. Chen et al.~\cite{chen2020arterylabel} use a GNN with hierarchical refinement for intracranial artery labelling on 3D vascular graphs, and Xie et al.~\cite{xie2024airway} combine CNN features with a position-aware GNN for airway-branch labeling, demonstrating that hybrid ``image backbone + graph reasoning'' pipelines are effective for fine-grained labeling of tubular structures. We extend this paradigm to a new clinical task --- node-level collateral detection on 2D DSA --- and introduce a shared node-probability space enabling fusion with pixel-level predictions.

\section{Dataset and Annotation}

\subsection{MAGIC Dataset}
Our primary data are drawn from the MAGIC repository, which consolidates multicentre clinical and radiological data from patients with acute ischemic stroke (AIS)~\cite{Baazaoui_2025}. From this repository, we select pre-interventional DSA sequences acquired on Siemens angiography systems from patients with anterior-circulation large-vessel occlusion stroke undergoing mechanical thrombectomy.
Each patient has two DSA views: an anterior--posterior (AP) projection and a lateral projection. Leptomeningeal collaterals are more readily identifiable in the AP view due to reduced vessel overlap; therefore, as a proof-of-concept, we restrict this work to AP frames. From each patient's full AP sequence, only frames in which collateral flow is visible were selected and annotated, yielding 136~AP collateral frames from 73~patients.

\subsection{DIAS Dataset}
To obtain vessel segmentation supervision, we leverage the public DIAS dataset~\cite{liu2024dias}, which provides 60 DSA sequences with expert-annotated intracranial artery masks at a $800\!\times\!800$ spatial-temporal resolution.
During inspection, we identified collateral vessels absent from the original DIAS ground truth; these were manually delineated and added to the training set to improve coverage of fine collateral vasculature.
DIAS is used exclusively to train the vessel segmentation model, which learns vascular appearance from DIAS and is then applied to MAGIC without requiring vessel annotations on the target domain.

\subsection{Annotation Workflow}\label{sec:annotation}
\begin{figure}[t]
  \centering
  \includegraphics[width=\textwidth]{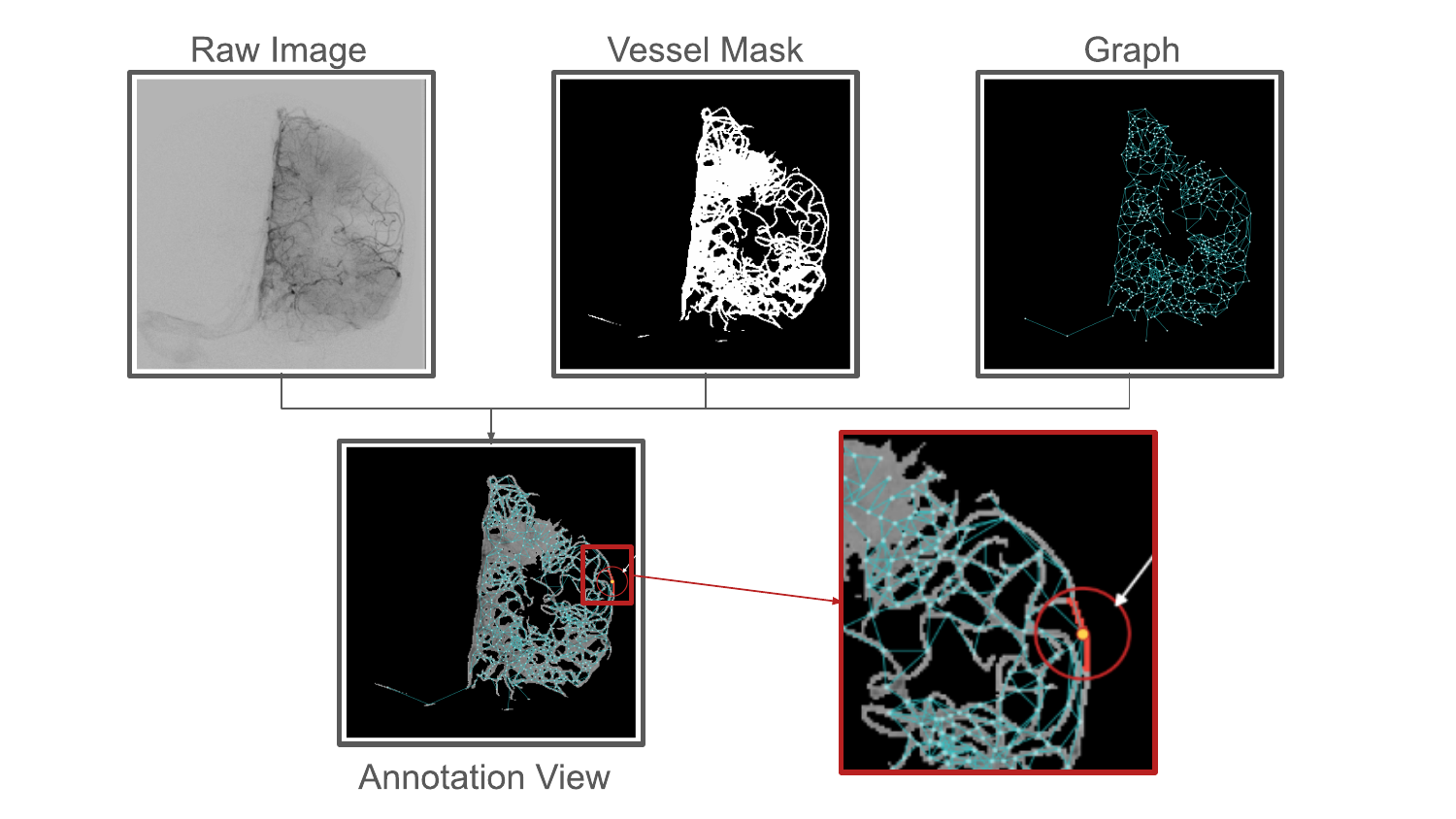}
  \caption{Data preparation and annotation workflow. \textbf{Top:} raw DSA frame, vessel mask, and derived vessel-segment graph. \textbf{Bottom:} the graph is overlaid on the DSA frame; a clinician clicks on a collateral segment to label the corresponding node. The zoom-in shows the pixel region associated with the selected node.}
  \label{fig:annotation}
\end{figure}

Collateral annotations are performed at the \emph{vessel-segment} level rather than by pixel-wise mask drawing.
For each DSA frame, a vessel mask is generated using the DIAS-trained segmentation model (Sec.~\ref{sec:vessel_mask}) and converted into a vessel-segment graph where each node represents one contiguous segment (Sec.~\ref{sec:graph_construction}). The graph is overlaid on the original DSA frame, and a neurology resident specializing in stroke inspects the overlay and selects nodes corresponding to collateral vessel segments, setting $\texttt{collateral}=1$ for each selected node.
This procedure is efficient, requiring only one click per collateral segment to produce an anatomically meaningful label tied directly to the graph representation used for classification. Since each graph node stores the pixel coordinates of its corresponding vessel segment (Sec.~\ref{sec:graph_construction}), pixel-level collateral masks can be directly derived from the node labels and serve as ground truth for both the pixel branch and baseline evaluation. Figure~\ref{fig:annotation} illustrates the annotation interface on a representative case.

\section{Method}

\subsection{Overall Framework}
\begin{figure}[t]
  \centering
  \includegraphics[width=\textwidth]{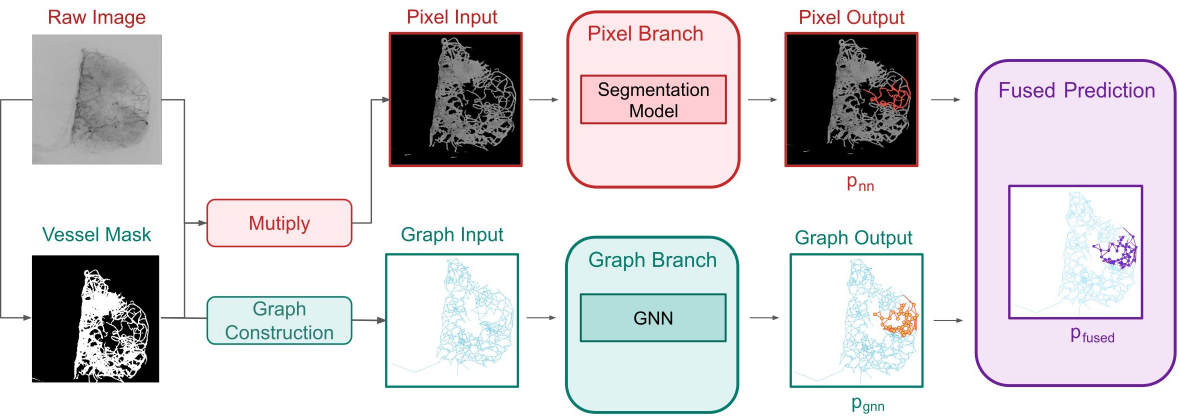}
  \caption{Overview of the hybrid graph-pixel framework. The \emph{pixel branch} (top) predicts dense collateral probabilities from the masked vessel image. The \emph{graph branch} (bottom) encodes each node with visual features and performs GNN message passing. Both outputs are fused at the node level.}
  \label{fig:pipeline}
\end{figure}

Given a single DSA frame, the pipeline first generates a binary vessel mask and converts it into a vessel-segment graph (Sec.~\ref{sec:vessel_mask}).
The framework comprises two complementary branches. The \emph{pixel branch} operates on local spatial appearance, suited to detecting compact, locally coherent collateral regions. The \emph{graph branch} reasons over vascular topology, propagating information along vessel connectivity so that each segment is classified in the context of its anatomical neighborhood.
Both branches are then projected into a shared node-probability space and fused for the final node-level prediction (Fig.~\ref{fig:pipeline}).

\subsection{Structural Parsing}\label{sec:vessel_mask}

\subsubsection{Vessel Mask Generation}
Since the MAGIC dataset lacks dense vessel annotations, we train a 2D nnUNet~\cite{isensee2021nnunet} on DIAS for binary vessel segmentation, using a combined objective with a topology-preserving clDice term~\cite{shit2021cldice}:
\begin{equation}\label{eq:cldice}
  \mathcal{L} = (1-\alpha)\,(\mathcal{L}_{\text{Dice}} + \mathcal{L}_{\text{CE}}) + \alpha\,\mathcal{L}_{\text{clDice}}\,,
\end{equation}
where $\mathcal{L}_{\text{Dice}}$ is the soft Dice loss, $\mathcal{L}_{\text{CE}}$ the cross-entropy loss, $\mathcal{L}_{\text{clDice}}$ the centerline Dice loss, and $\alpha$ controls the topology-preservation strength. We set $\alpha=0.3$ empirically: while varying $\alpha$ had only a marginal effect on volumetric overlap (Dice/IoU), the clDice term qualitatively improves the connectivity and completeness of the segmented collateral network. Leptomeningeal collaterals are extremely thin, and a purely Dice/CE-trained baseline tends to produce fragmented masks, whereas the topology-preserving term yields more continuous segments better suited to vessel-graph construction.
At inference, a five-fold ensemble is applied frame-by-frame to the MAGIC DSA data, producing binary vessel masks without requiring any vessel annotations on the target domain. Averaged over the five cross-validation folds on DIAS, the segmentation model achieves a mean Dice of 0.789 and IoU of 0.653.
\subsubsection{Graph Construction}\label{sec:graph_construction}
In this block, each binary vessel mask is converted into a vessel-segment graph. First, the mask is skeletonized and morphologically thinned to a one-pixel-wide centerline~\cite{lee1994skeletonize}. Second, skeleton pixels with degree~$\neq 2$ in their 8-connected neighbourhood are identified as topological nodes (junctions or endpoints), and connected paths of degree-2 pixels between them define edges. Finally, every foreground pixel in the vessel mask is assigned to its nearest edge, partitioning the mask into contiguous vessel segments.

Following a \emph{line-graph} convention~\cite{vesselgraph2022}, each vessel \emph{segment} becomes a graph \emph{node}, and two nodes are connected by an edge if their corresponding segments share a skeleton junction.
Isolated nodes and disconnected components are iteratively bridged by inserting edges between their closest centroid pairs until the graph is fully connected, ensuring that GNN message passing can reach all vessel segments.
Each node stores its bounding box, centroid, the nearest centerline point (\texttt{center\_proj}), and the full set of pixel coordinates (\texttt{pixels}) --- the latter being essential for the pixel-to-node conversion used for branch fusion (Sec.~\ref{sec:fusion}).

Applying this construction to 136 anterior-posterior (AP) collateral frames from 73 patients in the MAGIC dataset yields 82{,}953 vessel-segment nodes. Of these, only 3{,}689 (4.4\%) are labeled as collateral-positive, reflecting a severe class imbalance that motivates the loss weighting and evaluation metrics described in the following sections. Across frames, collateral-positive nodes range from compact clusters to dispersed multi-component patterns. The cohort primarily consists of middle cerebral artery occlusions, with 43 (58.9\%) M1 and 26 (35.6\%) M2+ cases, alongside 4 (5.4\%) ACA and ICA/Carotid-T occlusions.

\subsection{Graph Branch for Node-Level Collateral Detection}\label{sec:model}
The graph branch operates directly on the vessel-segment graph. It takes the graph topology and per-node visual features as input, performs message passing to aggregate neighbourhood context, and outputs a collateral probability for each node.

\paragraph{Node feature construction.}
Given a single DSA frame, all pixels outside the vessel mask are zeroed out and the masked image is encoded by a pretrained vision backbone, producing a dense feature map.
For each graph node, a fixed-dimensional representation~$\mathbf{x}_i$ is extracted by applying ROI-Align (region-of-interest feature pooling)~\cite{he2017maskrcnn} to the backbone feature map using the node's bounding box as the region of interest.
This ROI feature is concatenated with the node's bounding box height and width, which encode segment size and help the network distinguish small collateral vessels from larger trunk segments.
The backbone architecture is detailed in Sec.~\ref{sec:exp_impl}.

\paragraph{Graph reasoning.}
Node features are propagated over the vessel graph using a graph neural network (GNN~\cite{kipf2017gcn,velickovic2018gat,xu2019gin}; architecture detailed in Sec.~\ref{sec:exp_impl}). A classification MLP then maps the updated embeddings to a single logit per node, and a sigmoid yields the per-node collateral probability.

\subsection{Pixel Branch and Fusion}\label{sec:fusion}
The pixel branch is trained independently from the graph branch. We train a 2D nnUNet~\cite{isensee2021nnunet} on MAGIC data for binary collateral segmentation at the pixel level, supervised by a standard Dice + cross-entropy loss. The segmentation target is a binary collateral mask derived from the node-level annotations: all pixels belonging to collateral-positive nodes are labelled as foreground (Sec.~\ref{sec:annotation}).

To integrate the two branches, we map the pixel branch's dense probability maps onto the graph structure. For each node $i$, using its constituent pixel set $\mathcal{P}_i$ from structural parsing (Sec.~\ref{sec:graph_construction}), the pixel-branch probability $p_{\text{nn}}^{(i)}$ is the mean predicted collateral probability over those pixels:
\begin{equation}\label{eq:pix2node}
  p_{\text{nn}}^{(i)} = \frac{1}{|\mathcal{P}_i|}\sum_{(x,y)\in\mathcal{P}_i} \mathrm{prob}(x,y)\,,
\end{equation}
where $\mathrm{prob}(x,y) \in [0,1]$ denotes the pixel-level collateral probability predicted by the nnUNet at spatial location $(x,y)$.
With both branches now operating in a shared node-probability space, we compute the final collateral prediction as a weighted average:
\begin{equation}\label{eq:fusion}
  p_{\text{fuse}}^{(i)} = \lambda\, p_{\text{gnn}}^{(i)} + (1-\lambda)\, p_{\text{nn}}^{(i)}\,,
\end{equation}
where $p_{\text{gnn}}^{(i)}$ is the direct output of the graph branch and $\lambda$ is the fusion weight.

\section{Experiments}

\subsection{Implementation Details}\label{sec:exp_impl}

The graph branch is trained as a binary node classifier with binary cross-entropy loss.
Given the severe class imbalance ($\sim$ 1:21), a positive-class weight (empirically set to 10) is applied, since higher weights were found to over-correct, producing excessive false positives.
The vision backbone used for node feature extraction is DINOv3~\cite{simeoni2025dinov3}, with ROI-Align pooling over each node's bounding box.
Unless otherwise stated, the graph branch is trained for 100 epochs with Adam, a learning rate of $2\times 10^{-4}$, hidden dimension 128, and two graph layers.
Graph-branch augmentation comprises random rotations and flips (with node bounding boxes and pixel coordinates transformed accordingly) and edge dropout (rate 0.1) on the graph topology.
The pixel branch is implemented with 2D nnUNet and trained independently on aligned patient-level folds; nnUNet applies its own self-configured augmentation pipeline internally.
For fusion, Eq.~\ref{eq:fusion} is evaluated with a fixed weight $\lambda=0.77$, selected by a simple out-of-fold sweep over $\lambda \in [0,1]$ using node-level PR-AUC as the objective.
Training uses automatic mixed precision on a single NVIDIA A100 GPU.

\subsection{Evaluation Protocol}\label{sec:exp_protocol}
We perform patient-level five-fold cross-validation on the 136 selected AP collateral frames. In each fold, four subsets are used for training and model selection, while the fifth is held out exclusively for inference. This patient-level partitioning ensures that no images from the same individual appear in both the training and test sets. 

\subsection{Evaluation Metrics}\label{sec:exp_metrics}
Performance is evaluated at the node level, with each node representing an individual vessel segment. Given the severe class imbalance (4.4\% collateral-positive nodes), we adopt the area under the precision-recall curve (PR-AUC) as the primary metric, alongside ROC-AUC. For threshold-dependent evaluation, we report precision, recall, and F1-score at an operating point $\tau^*$ derived from the validation set. Following \cite{fernandez2005comparison}, this point is selected by minimizing the Euclidean distance to the ideal $\{1,1,1,1\}$ coordinate in the four-dimensional space of sensitivity (Se), specificity (Sp), positive predictive value (PPV), and negative predictive value (NPV):
\begin{equation}\label{eq:auto_thresh}
  \tau^{*} = \arg\min_{\tau} \sqrt{(1-\mathrm{Se})^2 + (1-\mathrm{Sp})^2 + (1-\mathrm{PPV})^2 + (1-\mathrm{NPV})^2}.
\end{equation}

\subsection{Baseline Models}\label{sec:baselines}
We compare the unified model against its individual branches and alternative architectures within each.
For the graph branch, we evaluate GCN~\cite{kipf2017gcn}, GAT~\cite{velickovic2018gat}, and GIN~\cite{xu2019gin}, all sharing the same node features and training protocol.
For the pixel branch, we compare nnUNet~\cite{isensee2021nnunet}, UNet~\cite{ronneberger2015unet}, and a DINOv3~\cite{simeoni2025dinov3} model that trains a Mask2Former~\cite{cheng2022mask2former} segmentation head on top of a frozen DINOv3 backbone. This DINOv3+Mask2Former variant already provides a transformer-based segmentation baseline; a broader comparison against other transformer architectures is left for future work.
All pixel baselines are trained for binary collateral segmentation using the node-derived pixel masks described in Sec.~\ref{sec:annotation}, and their dense probability maps are projected to node space via Eq.~\ref{eq:pix2node} for node-level evaluation.

\section{Results}

\subsection{Overall Node-Level Performance}
\begin{figure}[!b]
  \centering
  \includegraphics[width=\textwidth]{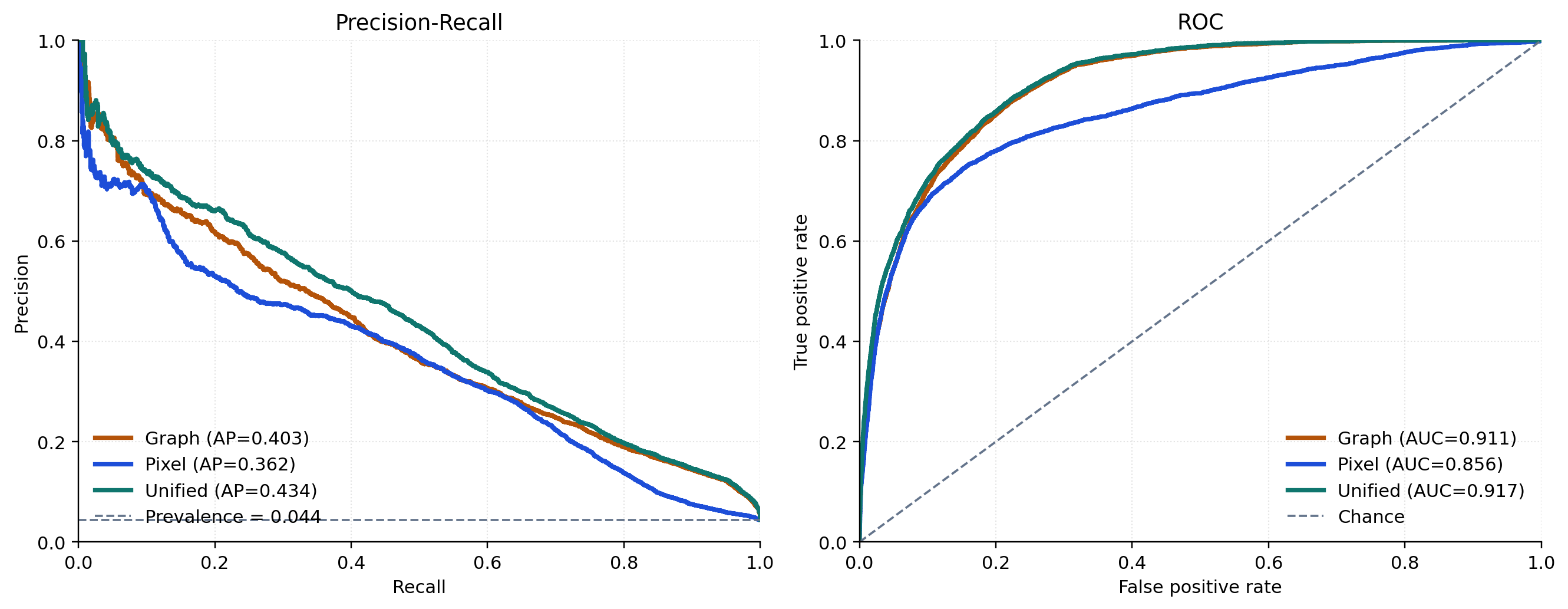}
  \caption{Node-level out-of-fold precision--recall and ROC curves for the graph branch, pixel branch, and unified (i.e., our proposed) model.}
  \label{fig:pr_roc}
\end{figure}

\begin{figure}[!b]
  \centering
  \begin{subfigure}[b]{0.48\textwidth}
    \centering
    \includegraphics[width=\textwidth]{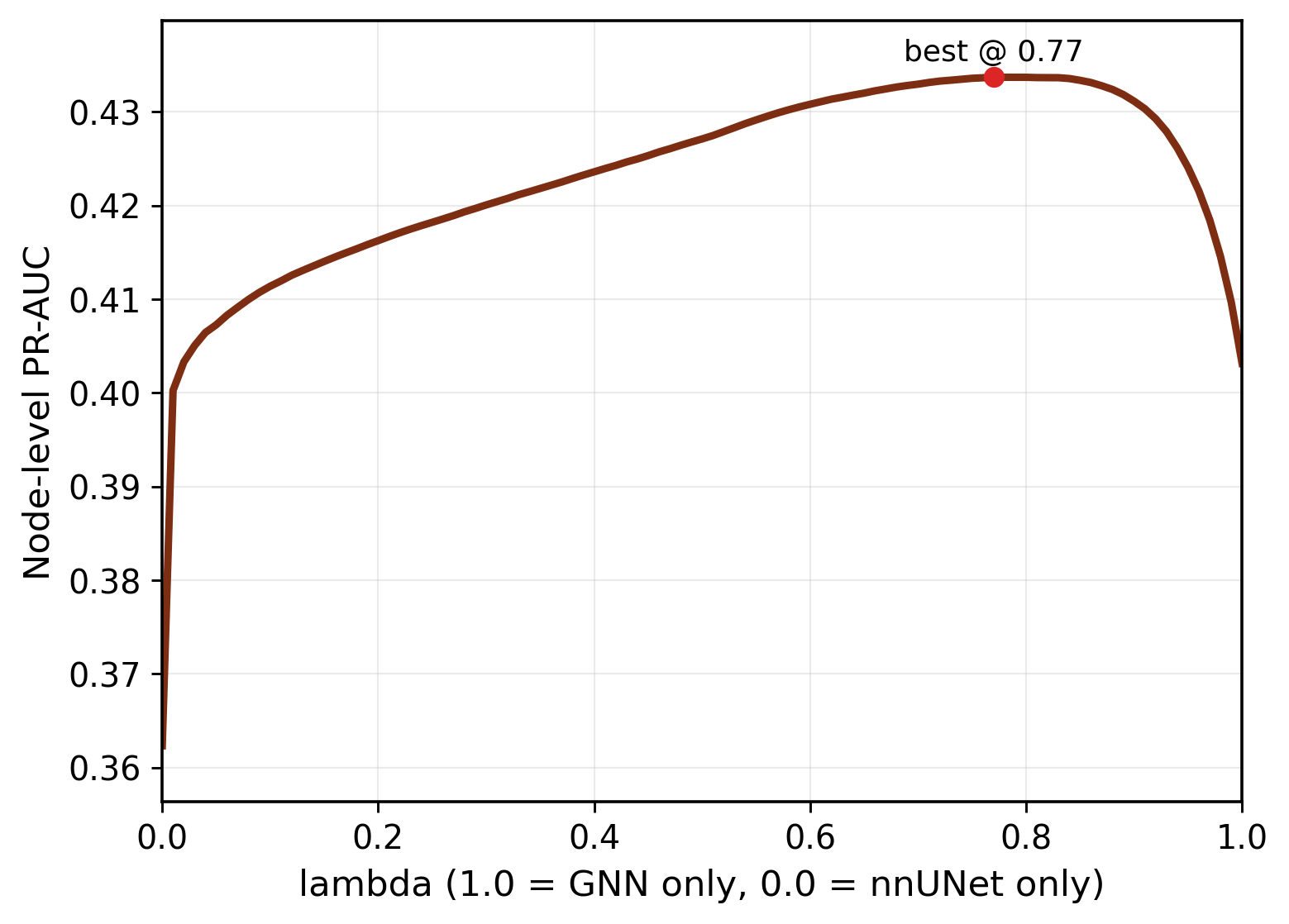}
    \caption{Fusion weight sweep.}
    \label{fig:lambda_sweep}
  \end{subfigure}
  \hfill
  \begin{subfigure}[b]{0.48\textwidth}
    \centering
    \includegraphics[width=\textwidth]{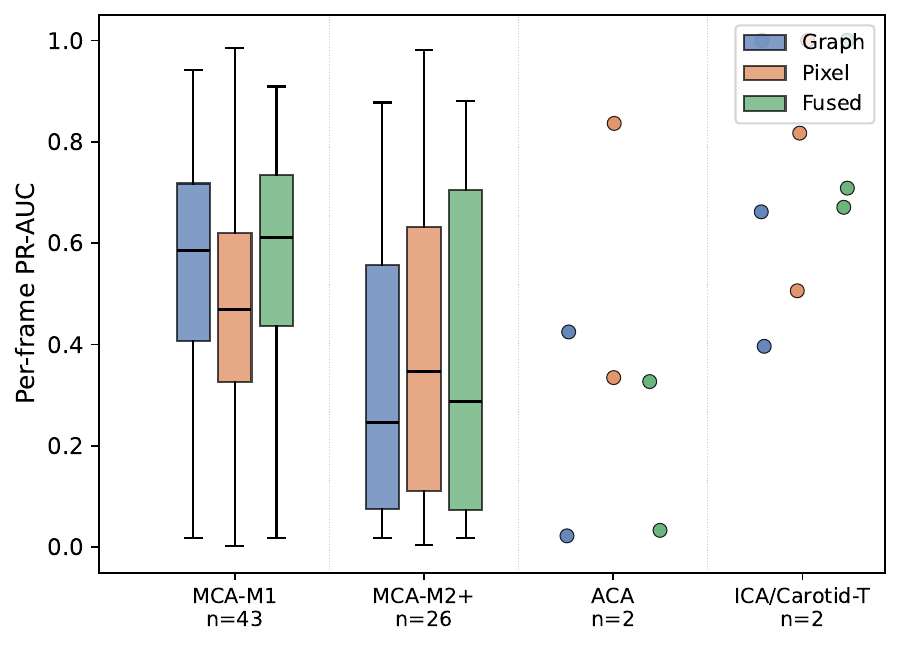}
    \caption{Performance by occlusion site.}
    \label{fig:occlusion_site}
  \end{subfigure}
  \caption{(a)~Node-level PR-AUC as a function of the fusion weight $\lambda$ ($\lambda{=}1$: graph only, $\lambda{=}0$: pixel only; optimum at $\lambda{=}0.77$). (b)~Per-frame PR-AUC by occlusion site; box plots for MCA-M1 and MCA-M2+, scatter points for ACA and ICA/Carotid-T (2 patients each).}
  \label{fig:lambda_occlusion}
\end{figure}

Table~\ref{tab:main_results} summarizes the LMC detection performance across the diverse models.
The proposed pixel-graph \emph{unified} model achieves the best node-level PR-AUC at 0.434, compared with 0.403 for the graph branch and 0.362 for the pixel branch.
The unified model also achieves the highest ROC-AUC and F1. A simple out-of-fold sweep over $\lambda \in [0,1]$ identified $\lambda=0.77$ as the best setting in terms of node-level PR-AUC (Fig.~\ref{fig:lambda_sweep}). The precision--recall and ROC curves (Fig.~\ref{fig:pr_roc}) confirm these trends, with the unified model dominating in PR space.
When stratified by occlusion site (Fig.~\ref{fig:occlusion_site}), the fused model consistently achieves the highest PR-AUC across the two major groups: MCA-M1 (0.570) and MCA-M2+ (0.385).
Absolute F1 remains modest across all models: under the severe 1:21 imbalance, F1 at a single operating point is threshold-sensitive and penalised by node-level false positives, whereas PR-AUC---integrating over all thresholds---is the more faithful summary and our primary metric.

\begin{table}[t]
  \centering
  \caption{Node-level out-of-fold performance. Precision, recall, and F1 are reported at the automatically selected operating point for each model.}
  \label{tab:main_results}
  \small
  \begin{tabular}{llccccc}
    \toprule
    Branch & Model & PR-AUC & ROC-AUC & Prec. & Recall & F1 \\
    \midrule
    \multirow{3}{*}{Graph}
      & GIN~\cite{xu2019gin} & 0.379 & 0.908 & 0.303 & 0.604 & 0.404 \\
      & GCN~\cite{kipf2017gcn} & 0.384 & 0.908 & 0.333 & 0.545 & 0.413 \\
      & GAT~\cite{velickovic2018gat} & 0.403 & 0.911 & 0.305 & 0.607 & 0.406 \\
    \midrule
    \multirow{3}{*}{Pixel}
      & UNet~\cite{ronneberger2015unet} & 0.292 & 0.887 & 0.303 & 0.501 & 0.377 \\
      & DINOv3~\cite{simeoni2025dinov3} & 0.278 & 0.789 & 0.348 & 0.470 & 0.400 \\
      & nnUNet~\cite{isensee2021nnunet} & 0.362 & 0.856 & 0.296 & \textbf{0.616} & 0.400 \\
    \midrule
    Unified & GAT~\cite{velickovic2018gat} + nnUNet~\cite{isensee2021nnunet} & \textbf{0.434} & \textbf{0.917} & \textbf{0.429} & 0.502 & \textbf{0.463} \\
    \bottomrule
  \end{tabular}
\end{table}

\subsection{Qualitative Analysis}
\begin{figure}[t]
  \centering
  \includegraphics[width=\textwidth]{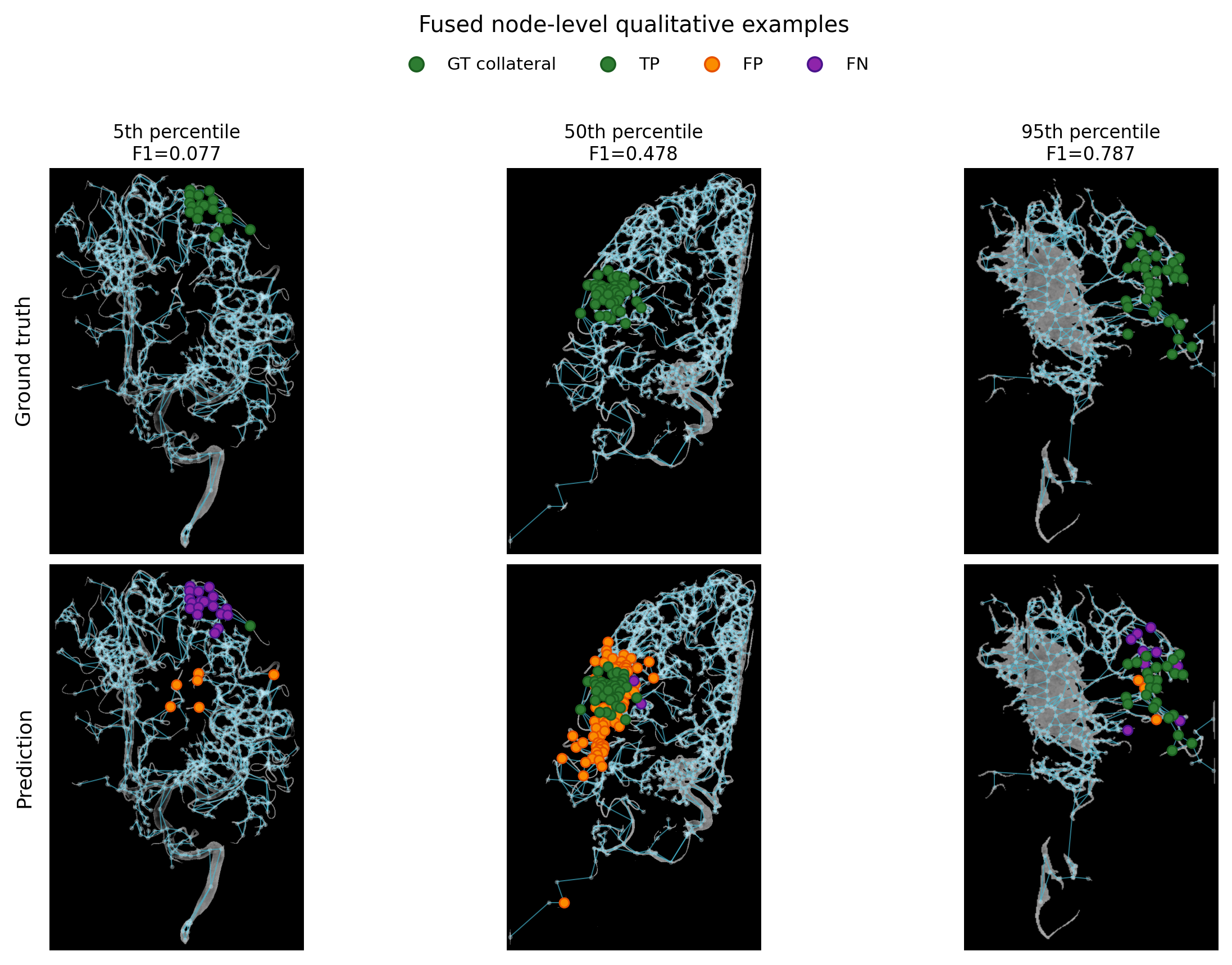}
  \caption{Qualitative results at the 5th (Low), 50th (Median), and 95th (High) percentiles of per-frame F1. Top: ground-truth collateral nodes (green). Bottom: predictions with true positives (green), false positives (orange), and false negatives (purple).}
  \label{fig:qualitative}
\end{figure}

Figure~\ref{fig:qualitative} presents qualitative node-level predictions of the fused model on three representative cases, selected at the 5th, 50th, and 95th percentiles of per-frame F1 score.
These cases reflect the diverse collateral configurations present in the dataset (Sec.~\ref{sec:graph_construction}), motivating a systematic analysis of how collateral topology affects model performance.

\subsection{Effect of Collateral Topology}
\begin{figure}[t]
  \centering
  \includegraphics[width=\textwidth]{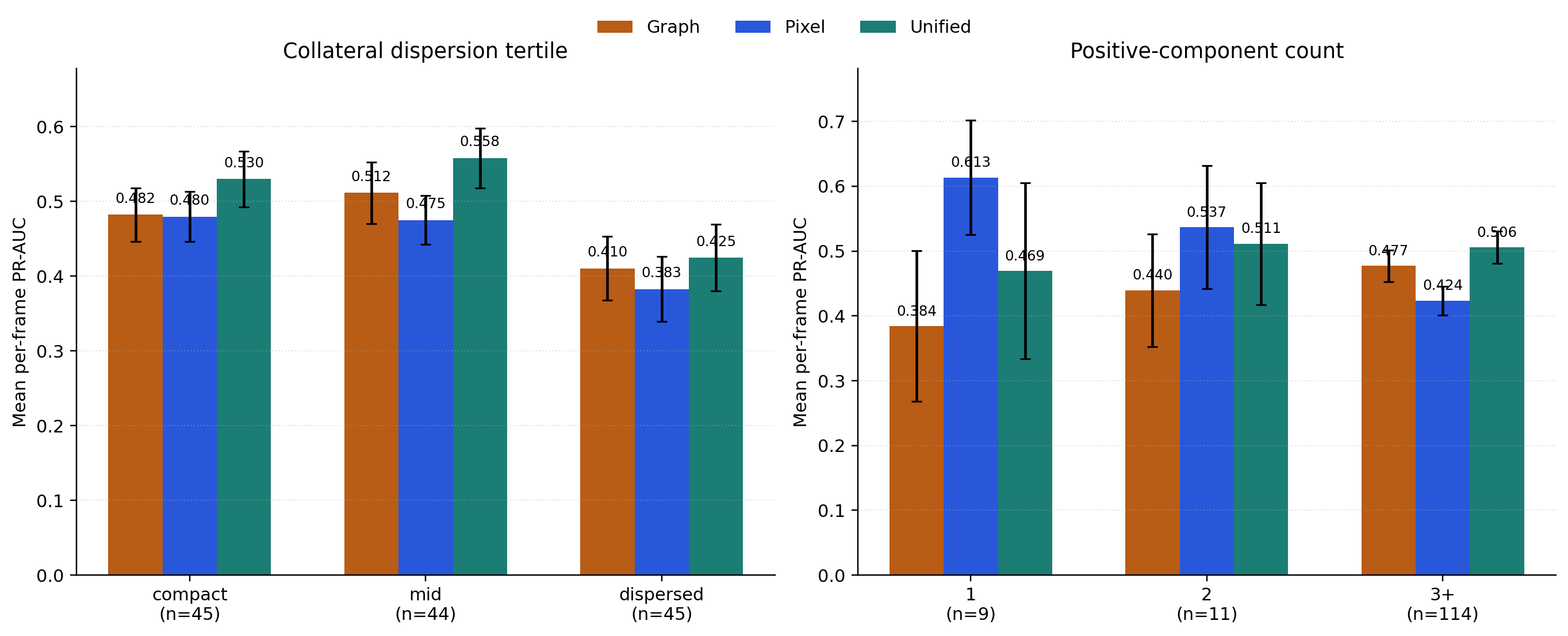}
  \caption{Mean per-frame PR-AUC by (a)~collateral dispersion tertile and (b)~number of connected components formed by collateral-positive nodes. Error bars: standard error of the mean.}
  \label{fig:topology}
\end{figure}

To examine when graph reasoning is most beneficial, we characterise collateral topology in two ways: by the spatial dispersion of collateral-positive nodes (Fig.~\ref{fig:topology}, left) and by the number of connected components they induce on the vessel graph (Fig.~\ref{fig:topology}, right). Both indicators reflect the degree to which collateral patterns are spatially scattered, requiring long-range relational reasoning that message passing is well-suited to capture.
As shown in Fig.~\ref{fig:topology}~(left), the graph branch maintains stable performance across all dispersion levels, with its advantage becoming most pronounced on dispersed collateral patterns.
The component-count view (Fig.~\ref{fig:topology}, right) tells a consistent story: graph-based reasoning is particularly beneficial when collateral is distributed across three or more disconnected components.

\subsection{Ablation Studies}

\paragraph{Node feature ablation.}
Table~\ref{tab:ablation} evaluates the effect of different node feature configurations on the GAT-based graph branch. The ROI feature captures local appearance from the vision backbone, while the height and width (HW) descriptors provide a geometric prior on vessel segment size.
Adding HW improves PR-AUC from 0.395 to 0.403 and increases recall (0.560~$\to$~0.607), indicating the size prior helps retrieve more collateral-positive nodes, which are typically smaller-radii segments. This recall gain comes at a slight precision decrease (0.340~$\to$~0.305), as some small non-collateral segments are also promoted, consistent with collateral vessels being finer than trunk segments.

\paragraph{Training strategy ablation.}
Table~\ref{tab:ablation} evaluates the effect of spatial data augmentation (random rotations and flips) on the graph branch. We ablate augmentation only for the graph branch, as nnUNet applies its own self-configured augmentation pipeline internally.
Under the ablation, PR-AUC --- our primary and threshold-independent metric --- decreases from 0.403 to 0.386, confirming that augmentation serves as an effective regulariser given the limited dataset size.


\begin{table}[!ht]
  \centering
  \caption{Ablation studies on the GAT-based graph branch.}\label{tab:ablation}
  \small
  \begin{tabular}{lccccc}
    \toprule
    Setting & PR-AUC & ROC-AUC & Prec. & Recall & F1 \\
    \midrule
    Proposed (ROI + HW, aug.)       & \textbf{0.403} & 0.911 & 0.305 & \textbf{0.607} & 0.406 \\
    \quad w/o HW features           & 0.395 & 0.908 & \textbf{0.340} & 0.560 & \textbf{0.423} \\
    \quad w/o augmentation          & 0.386 & \textbf{0.917} & 0.325 & 0.603 & 0.422 \\
    \bottomrule
  \end{tabular}
\end{table}
\section{Conclusion}

We present the first proof-of-concept study for automated identification of individual leptomeningeal collaterals on DSA, moving beyond subjective and ordinal grading towards objective, per-vessel characterization. By formulating the task as a node-level classification on a vessel-segment graph, our hybrid graph-pixel framework combines topology-aware reasoning with dense local evidence in a shared node-probability space. Our analysis shows the two schemes are complementary: the graph branch excels on spatially dispersed collateral patterns, while the pixel branch is more effective for compact, dense regions. Their fusion yields the best-performing model for individual collateral identification.

\subsubsection{Limitations and Future Work}
As a first step, this work operates on pre-selected DSA frames; an upstream frame-level selector, extension from AP to lateral views via multi-view learning, and use of the full temporal series for flow-dynamic cues are natural next stages. Annotations come from a single stroke neurologist, so inter-rater agreement was not assessed. The segmentation model is trained on DIAS and transferred to MAGIC, which lacks vessel ground truth; target-domain segmentation quality therefore cannot be quantified, and detection is bounded by the predicted vessel mask. The graph is made fully connected by bridging isolated components with nearest-centroid edges, which are purely geometric and may link physiologically unrelated segments; anatomically informed connectivity, e.g.\ geodesic minimum-spanning-tree reconstruction~\cite{moriconi2019vtrails}, is a promising alternative. Finally, evaluation is at the node and frame level on a small, single-centre/single-vendor cohort (73 patients, one Siemens system) without full clinical characteristics, precluding external validation. Patient-level aggregation, failure-case analysis of node-level false positives, multi-rater annotation, and larger multi-centre, multi-vendor data are left for future work.


\begin{credits}
\subsubsection{Acknowledgments}
EDLR, CP, SS and BM are supported
by the Helmut Horten Foundation.

\subsubsection{\discintname}
The authors have no competing interests to declare that are relevant to the content of this article.
\end{credits}

\nocite{vesselgraph2022,isensee2021nnunet}
\bibliographystyle{splncs04}
\bibliography{references}

\end{document}